\documentclass[aps,prl,twocolumn]{revtex4}
\usepackage{epsfig}
\begin{document}

\title{Isospin-projected nuclear level densities
by the shell model Monte Carlo method}

\author{H. Nakada$^1$ and Y. Alhassid$^2$}

\affiliation{$^1$Department of Physics, Graduate School of Science,
Chiba University, Inage, Chiba 263-8522, Japan\\
$^2$Center for Theoretical Physics,
Yale University, New Haven, Connecticut 06520, U.S.A.}


\begin{abstract}
We have developed an efficient isospin projection method in the shell
model Monte Carlo approach for isospin-conserving Hamiltonians. For
isoscalar observables this projection method has the advantage of being
exact sample by sample. The isospin projection method allows us to take
into account the proper isospin dependence of the nuclear interaction, thus avoiding  a sign problem that such an interaction introduces in unprojected calculations.
We apply our method in the calculation of the isospin dependence of
level densities in the complete $pf+g_{9/2}$ shell.
We find that isospin-dependent corrections to the total level density are particularly important for $N \sim Z$ nuclei.
\end{abstract}

\pacs{21.10.Ma, 21.60.Cs, 21.60.Ka, 21.60.-n}

\maketitle

The level density is among the most important statistical properties of nuclei. It is required for the calculation of transition rates through Fermi's golden rule and of nuclear reaction rates through the Hauser-Feshbach theory. Applications in astrophysics include estimates of neutron and proton radiative capture rates in the $s$, $r$, and $rp$ processes~\cite{BBF57,RTK97}.

Most theoretical approaches to the level density are based on the Fermi gas model, i.e., Bethe's formula~\cite{Bethe37}. This formula describes the level density of a non-interacting many-fermion system in terms of the single-particle level density parameter $a$. Shell corrections and two-body correlations (e.g., pairing correlations) are taken into account empirically by shifting the ground-state energy by an amount $\Delta$. The resulting backshifted Bethe's formula (BBF) for the total level density at excitation energy $E_x$ is given by
\begin{equation}\label{BBF}
\rho(E_x) = {\pi^{1/2}\over{12}} a^{-1/4}(E_x-\Delta+t)^{-5/4}
\exp\!\left[2\sqrt{a(E_x-\Delta)}\right] \;,
\end{equation}
where $t$ represents the Lang-LeCouteur modification~\cite{LLC54} defined by $at^2-t=E_x-\Delta$.
The BBF describes well the experimental level densities of many nuclei
if the parameters $a$ and $\Delta$ are fitted individually for each nucleus~\cite{Dilg73}. Although their global systematics have been studied extensively, these fitted parameters exhibit a significant dependence on the nucleus under consideration.  Since nuclear level densities increase very rapidly with $E_x$, an accurate determination of $a$ and $\Delta$ is crucial for reliable calculations of level densities and reaction rates.

As the basic microscopic model of nuclear structure, the interacting shell model
takes into account both shell effects and correlations. This model has been successful in describing the low-lying states of many nuclei and is also an
attractive framework to study microscopically thermal and statistical
properties of nuclei.
However, the large dimensionality of the many-particle model space has limited the applicability of direct diagonalization methods of the shell model Hamiltonian in
medium- and heavy-mass nuclei. This limitation has been overcome using
the shell model Monte Carlo (SMMC) approach~\cite{SMMC,ADK94}. While the
SMMC method cannot provide a detailed spectrum of the many-particle Hamiltonian, it enables calculation of thermal and statistical properties in model spaces that are many orders of magnitude larger than those that can be treated by conventional methods.

We have developed a method~\cite{NA97} to calculate level densities in the SMMC approach, and applied it to nuclei in the mass range $A \sim 50 -70$. We found good agreement with experimental data without any adjustable parameters~\cite{NA98,ALN99}. Fermionic Monte Carlo methods often suffer from the so-called sign problem, which leads to large statistical errors and a breakdown of the method at low temperatures. However, the dominating collective components of the nuclear interaction~\cite{DZ96} usually have a good sign, and general effective interactions containing small terms that have a bad sign can be treated with the method introduced in Ref.~\cite{ADK94}. In the calculation of level densities the sign problem was circumvented by constructing a good-sign interaction in the $pf + g_{9/2}$ shell~\cite{NA97} that correctly includes the dominating collective components of realistic effective interactions.

In the SMMC approach, thermal observables are averaged over all possible values of the good quantum numbers. However, many of the applications require the dependence of thermal observables on the good quantum numbers. This dependence is determined in  the SMMC approach by projection methods. Parity projection was implemented in the study of the parity dependence of level densities~\cite{NA97,NA98,ABLN,Oz07}, and a spin projection method was introduced to determine the spin distribution of nuclear levels~\cite{ALN07}. The particle-number reprojection technique has been developed~\cite{ALN99} to facilitate the systematic studies of level densities for a large number of nuclei.

The isospin $T$ is approximately a conserved quantum number in nuclei. Reliable treatment of the isospin dependence of the level densities
could be important in $N \sim Z$ nuclei because levels with different $T$ values lie close in energy.
In our SMMC studies of level densities we have used a good-sign Hamiltonian
to keep the statistical errors small.
While this Hamiltonian contains the dominating collective components of
nuclear effective interactions, it might not describe properly the
energy differences of states with different isospin values.
Such discrepancies can be corrected by including an isospin-dependent interaction~\cite{Or97,La98}. However such an interaction has a bad Monte Carlo sign and was treated perturbatively in Refs.~\cite{Or97,La98}. Here we overcome this problem by introducing an exact isospin projection method in the SMMC approach. This $T$-projection enables us to determine the isospin dependence of nuclear levels and to include accurately an isospin-dependent interaction term. We apply the method to the calculation of level densities in the complete $pf+0g_{9/2}$ shell.

The SMMC method is based on the Hubbard-Stratonovich (HS) representation
of the imaginary-time propagator $e^{-\beta  H}$ describing the Gibbs ensemble of
a nucleus with Hamiltonian $H$ at inverse temperature $\beta$. In the HS decomposition $e^{-\beta  H} = \int D[\sigma] G_\sigma U_\sigma$, where
$G_\sigma$ is a Gaussian weight and $U_\sigma= \mathcal{T} e^{-\int_0^\beta d\tau h_\sigma(\tau)}$ is the propagator of the one-body Hamiltonian $h_\sigma$
describing non-interacting nucleons moving in time-dependent auxiliary fields $\sigma(\tau)$  ($\mathcal{T}$ denotes time ordering).
The canonical expectation value of an observable $O$ is then given by
\begin{equation}\label{observable}
\langle O \rangle_{A, T_z} = \left\langle {{\rm Tr}_{A, T_z}(O U_\sigma)
\over {\rm  Tr}_{A, T_z} U_\sigma} \right\rangle_W \;,
\end{equation}
where ${\rm Tr}_{A,T_z}$ denotes a trace
in the subspace of the fixed mass number $A$ and isospin component $T_z$. Since $A=N+Z$ and $T_z=(N-Z)/2$, this trace is equivalent to a trace at fixed neutron number $N$ and proton number $Z$. We have also used the notation
$\langle X_\sigma \rangle_W \equiv \int D[\sigma] W(\sigma) X_\sigma /
\int  D[\sigma] W(\sigma)$
with $W(\sigma)\equiv G_\sigma {\rm Tr}_{A,T_z}U_\sigma$. Here and in
the following we assume ${\rm Tr}_{A,T_z}U_\sigma \geq 0$ (i.e., we
assume a good sign interaction and an even-even or $N=Z$ nucleus).
Otherwise the sign function ${\rm  Tr}_{A, T_z} U_\sigma/|{\rm  Tr}_{A, T_z} U_\sigma|$ has to be included.
In the Monte Carlo method we choose $M$ samples of the $\sigma$ fields
according to the weight function $W(\sigma)$ (each sample denoted by $\sigma_k$), and estimate $\langle X_\sigma \rangle_W \approx \sum_k X_{\sigma_k} /M$.

Let us consider a model space for a set of isobars (i.e., nuclei with fixed $A$
but different values of $T_z$). Since the isospin forms an $su(2)$ algebra, isospin projection generally requires a three-dimensional integration
over the Euler angles in isospin space (in analogy with angular
momentum projection). However, this integration is time consuming and in
the following we describe a more efficient method. We assume the nuclear Hamiltonian $H$ to be isospin invariant (i.e., $[H,{\mathbf T}]=0$), in which case both $T$ and $T_z$ are good quantum numbers and the isospin multiplets (at fixed $T$) are degenerate with respect to $T_z$. The subspace with fixed $T_z$ (i.e., fixed nucleus) contains all energy eigenstates with $T\geq | T_z|$. In the following we assume $N\geq Z$ without loss of generality, and denote by $T_0=(N-Z)/2 \ge 0$ the value of $T_z$ for a specific nucleus. If the operator $X$ is an isoscalar, we can decompose its trace for the specific nucleus as
\begin{equation}\label{trace-Tz}
{\rm Tr}_{A,T_z=T_0} X = \sum_{T\geq T_0} {\rm Tr}_{A,T} X \;,
\end{equation}
where ${\rm Tr}_{A,T}$ represents the trace
for fixed $A$ and isospin $T$ (not including degeneracy with respect to $T_z$).
The trace of $X$ for a specific isospin $T=T_0$ is then obtained by
\begin{equation}\label{trace-T}
{\rm Tr}_{A, T=T_0} X = {\rm Tr}_{A,T_z=T_0} X
- {\rm Tr}_{A, T_z=T_0+1} X \,.
\end{equation}
We note that $T_z=T_0+1$ corresponds to a nucleus
with $(N+1)$ neutrons and $(Z-1)$ protons. In the level density calculations we use $X=He^{-\beta H}$~\cite{NA97}, which is an isoscalar.
If $X$ is not an isoscalar, a $T_z$-dependent factor
is required in Eq.~(\ref{trace-Tz}) and hence in Eq.~(\ref{trace-T}).

An isospin-invariant two-body interaction can always be
decomposed into a sum of squares of {\em isoscalar} one-body operators~\cite{SMMC}. Therefore , the one-body Hamiltonian $h_\sigma$ and the propagator $U_\sigma$ (in the corresponding HS representation) remains isospin invariant for any configuration of the $\sigma$ fields. Equation~(\ref{trace-T}) then holds with $X=OU_\sigma$ as long as  $O$ is an isoscalar observable
\begin{equation}\label{trace-T'}
{\rm Tr}_{A,T=T_0} (OU_\sigma) = {\rm Tr}_{A,T_z=T_0} (OU_\sigma)
- {\rm Tr}_{A,T_z=T_0+1} (OU_\sigma) \,.
\end{equation}
Thus, $T$ projection can be carried out for each sample by using the simpler $T_z$ projection. Equation (\ref{trace-T'}) guarantees that $T$ projection is implemented exactly {\em sample by sample}. An equation analogous to Eq.~(\ref{trace-T}) was
used for angular momentum projection with the scalar operator $X=O
e^{-\beta H}$~\cite{ALN07}.  However, in contrast to the isospin projection,  $h_\sigma$ in the HS transformation is not a scalar under spatial rotations so the equation analogous to (\ref{trace-T'}) does not hold
sample by sample. We therefore expect that the present $T$ projection method
leads to statistical errors that are typically smaller than the statistical errors in the angular momentum projection of Ref.~\cite{ALN07}.

Based on Eq.~(\ref{trace-T'}), isospin projection can be implemented using
the particle-number {\em reprojection} technique~\cite{ALN99},
in which the Monte Carlo sampling is done for a reference  nucleus $(N,Z)$ and thermal observables are evaluated by reprojection on a nucleus $(N',Z')$.
In particular, the ratio between the $T$- and $T_z$-projected partition
functions
is given by
\begin{eqnarray}\label{partition-ratio}
{Z_{A,T=T_0}(\beta) \over Z_{A,T_z=T_0}(\beta)} =
1-\left\langle { {\rm Tr}_{A,T_z=T_0+1} U_\sigma \over
{\rm Tr}_{A,T_z=T_0} U_\sigma} \right\rangle_W \;.
\end{eqnarray}
Similarly, the expectation value of an observable $O$
for isospin $T=T_0$ can be calculated from
\begin{widetext}
\begin{equation}\label{observable'}
\langle O \rangle_{A,T=T_0}= {\left\langle
{{\rm Tr}_{A,T_z=T_0}(OU_\sigma) \over
 {\rm Tr}_{A,T_z=T_0} U_\sigma }\right\rangle_W -
\left\langle
{{\rm Tr}_{A,T_z=T_0+1}(OU_\sigma) \over
 {\rm Tr}_{A,T_z=T_0+1} U_\sigma} \cdot
{{\rm Tr}_{A,T_z=T_0+1} U_\sigma \over
 {\rm Tr}_{A,T_z=T_0} U_\sigma} \right\rangle_W
\over 1 - \left\langle { {\rm Tr}_{A,T_z=T_0+1} U_\sigma \over
{\rm Tr}_{A,T_z=T_0} U_\sigma} \right\rangle_W} \;.
\end{equation}
\end{widetext}
The Monte Carlo sampling in Eqs.~(\ref{partition-ratio}) and (\ref{observable'}) is carried out according to the weight function $W(\sigma)$ for the nucleus $A, T_0$. Equation~(\ref{observable'}) (that includes reprojection on the nucleus $A,T_0 + 1$) is then used
to calculate $\langle O \rangle_{A,T=T_0}$.

Isospin projection on $T=T_0+1$ can be implemented within the same Monte Carlo sampling using
\begin{widetext}
\begin{equation}\label{observable''}
\langle O \rangle_{A,T=T_0+1}= {\left\langle
{{\rm Tr}_{A,T_z=T_0+1}(OU_\sigma) \over
 {\rm Tr}_{A,T_z=T_0+1}(U_\sigma)} \cdot
{{\rm Tr}_{A,T_z=T_0+1}(U_\sigma) \over
 {\rm Tr}_{A,T_z=T_0}(U_\sigma)}\right\rangle_W -
\left\langle
{{\rm Tr}_{A,T_z=T_0+2}(OU_\sigma) \over
 {\rm Tr}_{A,T_z=T_0+2}(U_\sigma)} \cdot
{{\rm Tr}_{A,T_z=T_0+2}(U_\sigma) \over
 {\rm Tr}_{A,T_z=T_0}(U_\sigma)} \right\rangle_W
\over \left\langle
{{\rm Tr}_{A,T_z=T_0+1}(U_\sigma) \over
 {\rm Tr}_{A,T_z=T_0}(U_\sigma)}\right\rangle_W -
 \left\langle { {\rm Tr}_{A,T_z=T_0+2}(U_\sigma) \over
{\rm Tr}_{A,T_z=T_0} U_\sigma} \right\rangle_W} \;.
\end{equation}
\end{widetext}
The projection on higher values of $T$ is carried out in a similar manner.
Thus we can project on all possible isospin values $T$ using the same Monte Carlo sampling.  In particular, we can calculate the level densities of isobars
with a single Monte Carlo sampling process as in the number reprojection method of Ref.~\cite{ALN99}.

The good-sign Hamiltonian of Ref.~\cite{NA97}, while reproducing the
 proper collective features at fixed isospin $T$, does not necessarily
 reproduce the correct isospin dependence of energy levels.
This isospin dependence can be particularly important for $N \sim Z$ (i.e., $T_z\sim 0$) nuclei, in which number of levels with $T=T_0=(N-Z)/2$ and  $T=T_0+1$ are comparable even close to the ground state.
A simple way to account for the proper isospin dependence of the nuclear interaction is to add an appropriate function of ${\mathbf T}^2$,
$f({\mathbf T}^2)$, to the effective Hamiltonian.
The simplest such function is $f({\mathbf T}^2)=\alpha{\mathbf T}^2$
($\alpha$ is a constant), as in the modified surface delta interaction~\cite{BG77}. This ${\mathbf T}^2$ term is repulsive ($\alpha >0$) and leads to a sign problem in the SMMC method. In Refs.~\cite{Or97,La98} such a $T$-dependent interaction term was treated perturbatively. However, the criterion for applicability of perturbation theory does not always hold.
As we explain below, the $T$ projection method enables us to account exactly for an arbitrary function $f({\mathbf T}^2)$ in the Hamiltonian.

To determine the proper isospin dependence in the Hamiltonian, we
extract from experimental data the excitation energy $E_x(T)$ of the lowest level for each isospin value $T$.
When not directly measured, the experimental values of $E_x(T)$ can be obtained
from the measured masses of the corresponding isobars, together with estimates of the Coulomb energy differences among them. For the latter we use $E_C(A,Z)-E_C(A,Z-1) =
{3\over 5}(2Z-1) e^2/R_C$ with $R_C=1.24A^{1/3}$\,fm.
We then shift the calculated value of $E(T)$ (the lowest energy of isospin $T$),
by $\delta E(T)=E(T_0)+E_x(T)-E(T)$ (for $T > T_0$),
so as to reproduce the experimental value of $E_x(T)$.
This shift defines the function $f(\mathbf{T}^2)$ in the effective Hamiltonian.
 The modified $T$-projected level densities are then determined by
 shifting the excitation energy at each $T$ by $\delta E(T)$.

We first apply the isospin projection method to the nucleus $^{58}$Cu. The effective interaction consists of
$T=1$ pairing interaction and surface-peaked multipole-multipole
interaction terms (quadrupole, octupole and hexadecupole) as in
Ref.~\cite{NA97}. Since $^{58}$Cu is an $N=Z$ (i.e., $T_0=0$) odd-odd
nucleus, its $T=1$ levels have analog levels in its neighboring
even-even nuclei with $T_z=\pm 1$. The lowest $T=0$ and $T=1$
levels can be quite close because of the pairing energy. The experimentally observed ground state of $^{58}$Cu has $T=0$, while the lowest $T=1$ state is close to it  with an excitation energy of $E_x(T=1)=0.203$\,MeV.
However, for the good-sign Hamiltonian of Ref.~\cite{NA97}, we find
$E(T=1)$ to be well below $E(T=0)$. 
The energy $E(T=1)$, describing the ground state of the even-even nucleus with $T_z=1$, is determined directly from $\langle H\rangle_{A,T_z=1}=\langle\mathrm{Tr}_{A,T_z=1}(HU_\sigma)/\mathrm{Tr}_{A,T_z=1}(U_\sigma)\rangle_W$ at low temperatures (as in Ref.~\cite{NA98}), while  $E(T=0)$ is determined from Eq.~(\ref{observable'}) with $T_0=0$ and $O=H$.
All other energies $E(T)$ (for $T> 1$) are determined from $\langle
H\rangle_{A,T_z=T}$, similarly to $E(T=1)$. Once we determine the energies $E(T)$, we adjust their differences to match the experimental values of $E_x(T)$.

\begin{figure}
\epsfysize= 0.85\columnwidth \centerline{\epsffile{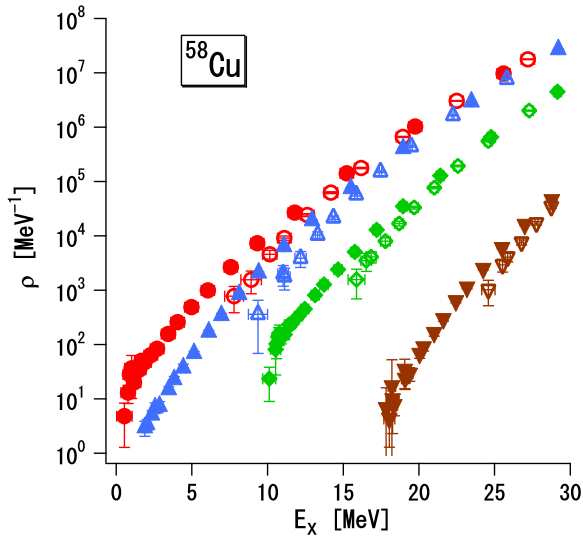}}
\caption{
$T\pi$-projected level densities of $^{58}$Cu,
obtained from SMMC calculations and shifted to match the experimental values of $E_x(T)$. Symbols are as follows: solid (open) circles for $T=0$, $\pi=+$ ($-$), solid (open) triangles for $T=1$, $\pi=+$ ($-$),
solid (open) diamonds for the $T=2$, $\pi=+$ ($-$),
and solid (open) inverted triangles for $T=3$, $\pi=+$ ($-$) densities.
}
\label{fig:Cu58t}

\end{figure}

In Fig.~\ref{fig:Cu58t} we present the $T$- and $\pi$-projected level densities
of $^{58}$Cu with the corrected values of $E(T)$.
An empirical BBF for the $T$-projected level densities (including the Lang-LeCouteur modification~\cite{LLC54}) is given by 
\begin{eqnarray}\label{BBF-T}
\rho_T(E_x)  = & g(2T+1){\pi^{5/2}\over{72}} a^{-3/4}(E_x-\Delta+t)^{-7/4} \nonumber \\
& \;\;\;\;\;\;\;\;\;\;\;\times\exp\!\left[2\sqrt{a(E_x-\Delta)}\right] \;,
\end{eqnarray}
where $g=1$ for a $T\pi$-projected density
and $g=2$ when parity is not projected.
The projected SMMC level densities can be well fitted to Eq.~(\ref{BBF-T})
if we use $T\pi$-dependent values for $a$ and $\Delta$.
The correction to $E(T)$ is described by a shift of the corresponding $\Delta$, but does not affect $a$.

\begin{figure}
\epsfysize= 0.8\columnwidth \centerline{\epsffile{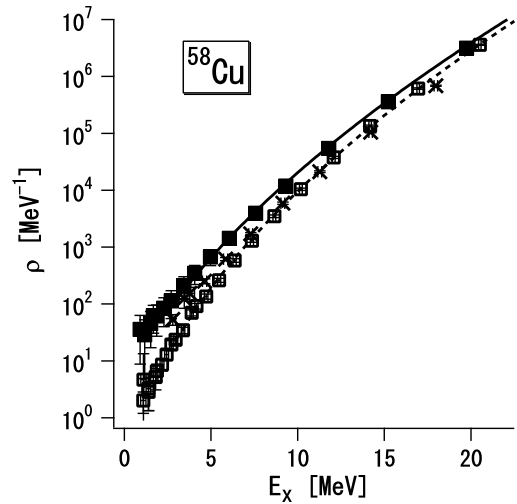}}
\caption{Total SMMC level densities of $^{58}$Cu.
Solid squares represent the density
obtained from the sum of the $T\pi$-projected densities
after the lowest energy for each isospin $T$ is corrected by $\delta E(T)$, while the open squares describe the total density
calculated from unprojected SMMC. Lines are fit to the BBF (\ref{BBF}).
Crosses are the results of the perturbative $\alpha{\mathbf T}^2$ correction
with $\alpha=1.2$\,MeV~\protect\cite{La98}.
}
\label{fig:Cu58n}

\end{figure}

The total level density is obtained by summing over the $T\pi$-projected level densities (after taking into account the shifts $\delta E(T)$).
The corresponding total level density for  $^{58}$Cu is shown in
Fig.~\ref{fig:Cu58n} (solid squares). For comparison we also show the
total level density obtained by a direct SMMC calculation (i.e., without
$T$ projection) for the interaction in Ref.~\cite{NA97} (open
squares). Fitting the SMMC density to Eq.~(\ref{BBF}),
we obtain $a=5.550\pm 0.021\,\mathrm{MeV}^{-1}$ and $\Delta=-0.658\pm
0.076\,\mathrm{MeV}$ after projection and $\delta E(T)$ correction,
while $a=5.827\pm 0.022\,\mathrm{MeV}^{-1}$ and $\Delta=1.069\pm
0.078\,\mathrm{MeV}$ for the unprojected density.
We observe that the inclusion of the proper isospin dependence in the Hamiltonian leads to a significant enhancement of the total level density at low excitation energies. This can be understood as follows. In the unprojected SMMC calculation, the ground state has $T=1$ and the level density at low energies is dominated by its $T=1$ component. The pairing energy in the $T=1$ component leads to a large $\Delta$ and thus too low level density at low energies. Once we include the proper isospin dependence (using the $T$ projection method), the ground state has $T=0$ and the level density at low energies is enhanced.

We also show in Fig.~\ref{fig:Cu58n} the total density obtained
from the perturbative $\alpha{\mathbf T}^2$ correction~\cite{Or97,La98}(crosses).
In this approach, the energy shifts of the $T>T_0(=0)$ components are assumed to be small. We find that the total level density enhancement at low excitation energies to be too small in the perturbative approach.

\begin{figure}[t]

\epsfysize= 0.85\columnwidth \centerline{\epsffile{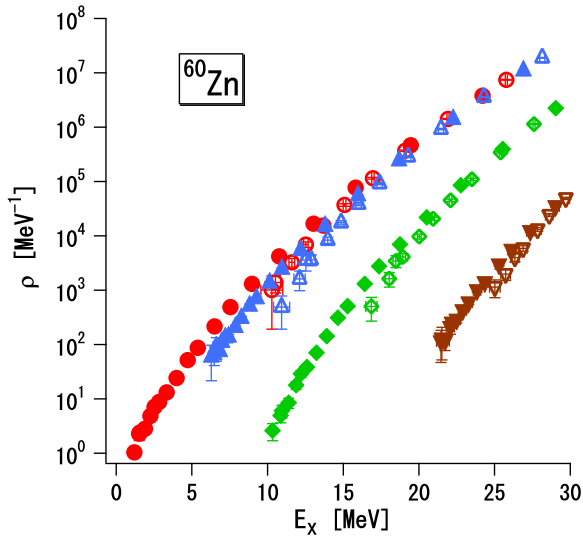}}
\caption{
$T\pi$-projected level densities of $^{60}$Zn
obtained in the SMMC method. Symbols and lines are as in
Fig.~\protect\ref{fig:Cu58t}.
}
\label{fig:Zn60t}
\end{figure}

\begin{figure}[t!]
\epsfysize= 0.8\columnwidth \centerline{\epsffile{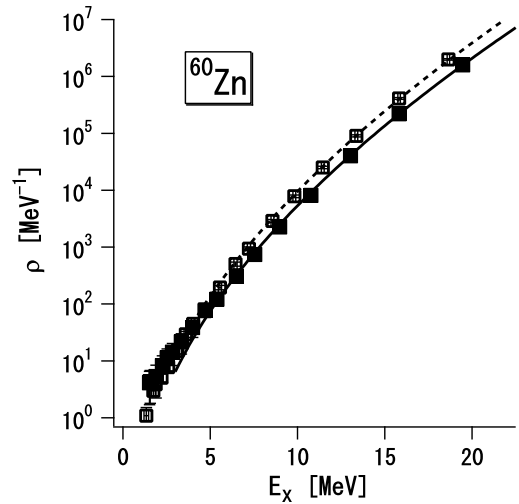}}
\caption{
SMMC total level densities of $^{60}$Zn.
Symbols are as in Fig.~\protect\ref{fig:Cu58n},
except that the perturbative result is not shown here.
}
\label{fig:Zn60n}
\end{figure}

Figure \ref{fig:Zn60t} shows the $T$- and $\pi$-projected level densities
of the even-even $N=Z$ nucleus $^{60}$Zn after correcting the values of $E(T)$.
In Fig.~\ref{fig:Zn60n} we compare the total density obtained from the
sum of the projected densities with the total level density determined
from a direct SMMC calculation (with no $\delta E(T)$ shifts). A fit to Eq.~(\ref{BBF}) gives $a=5.947\pm
0.020\,\mathrm{MeV}^{-1}$ and $\Delta=2.102\pm 0.053\,\mathrm{MeV}$
for the corrected density ($a=6.185\pm 0.009\,\mathrm{MeV}^{-1}$ and
$\Delta=1.679\pm 0.059\,\mathrm{MeV}$ for the unprojected density).
In this nucleus the ground state has $T=0$ both experimentally and in the unprojected SMMC calculation. Hence the level density at low energies is dominated by its $T=0$ component either with or without the $\delta E(T)$ corrections.
Nevertheless, we observe that the $\delta E(T)$ corrections are
important at higher energies $E_x\agt 10$\,MeV, where they suppress the
total level density. To explain this, we note that there are two
counteracting effects in the isospin dependence of the level density:
the $(2T+1)$ factor in Eq.~(\ref{BBF-T}) tends to enhance the projected
density for larger values of $T$ while the increase of  $\Delta$ with
$T$ tends to suppress the projected density at larger $T$. We find that
the $T=1$ component is dominant at $E_x\agt 10$\,MeV
in the unprojected total density. In contrast, once we take
into account the proper energy shift $\delta E(T=1)\approx 1.5$\,MeV, we find the $T=1$ density to be substantially suppressed and comparable to the $T=0$ density for $E_x\agt 10$\,MeV, as is seen in Fig.~\ref{fig:Zn60t}. This results in an overall suppression of the total level density for the higher energy regime.  For this nucleus the perturbative method gives almost the same level density as the unprojected density.

While the isospin-dependent corrections are important in $T_z=0$ nuclei
(and possibly in some $|T_z|=1$ odd-odd nuclei), we have confirmed that these corrections are insignificant for $|T_z|>0$ nuclei (at least up to $E_x\sim 20$\,MeV). Since levels with $T>T_0$ ($T_0 \geq 1$) are well separated in energy
from the $T=T_0$ levels already with the good-sign Hamiltonian,
the density of all $T>T_0$ levels is less than half the density of the $T=T_0$ levels.

In summary, we have developed an efficient isospin projection method
for an isospin-conserving Hamiltonian in the SMMC approach and applied
it to nuclei in the complete $pf+g_{9/2}$ shell. This isospin projection
is exact sample by sample and thus leads to statistical errors that are typically smaller than the statistical errors in a corresponding spin projection method. We have used this projection method to take into account the proper isospin dependence of the nuclear interaction, avoiding a sign problem that occurs when such an isospin-dependent interaction is included in unprojected calculations. For $N=Z$ nuclei, we find that this isospin dependence can lead to significant corrections to the total level density.

This work is supported in part by the U.S. DOE grant No. DE-FG-0291-ER-40608
and as Grant-in-Aid for Scientific Research (C), No.~19540262, by the MEXT,
Japan. Computations were carried out on the PC cluster Helios and IBM SP3 in JAERI, and on CP-PACS at the Center for Computational Physics at the
University of Tsukuba.

\end{document}